\documentclass[draft,jgrga]{agutex}

\usepackage[dvips]{graphicx}
\usepackage[toc,page]{appendix}
\usepackage{graphicx}
\usepackage{amsmath}
\usepackage[varg]{txfonts}
\usepackage[usenames]{color}
\usepackage{xspace}
\usepackage{siunitx}
\usepackage{url}
\usepackage{longtable}

\usepackage[dvips]{graphicx}

\usepackage{adjustbox}

\usepackage{color}

\begin{document}

\title{Analysis of the spatial non-uniformity of the electric field in spectroscopic diagnostic methods of atmospheric electricity phenomena}

\authorrunninghead{MALAG\'ON-ROMERO et al.}

\titlerunninghead{Spatial non-uniformity of air discharges}

\authors{
  A. Malag\'on-Romero \altaffilmark{1},
  F. J. P\'erez-Invern\'on \altaffilmark{1},
 A. Luque \altaffilmark{1},
 F.J. Gordillo-V\'azquez \altaffilmark{1}.
}

\affil{1}{Instituto de Astrof\'isica de Andaluc\'ia (IAA), CSIC, PO Box 3004, 18080 Granada, Spain}


\authoraddr{A. Malag\'on-Romero,
Instituto de Astrof\'isica de Andaluc\'ia (IAA),
CSIC, PO Box 3004, 18080 Granada, Spain. (amaro@iaa.es)
}

\authoraddr{F. J. P\'erez-Invern\'on,
Instituto de Astrof\'isica de Andaluc\'ia (IAA),
CSIC, PO Box 3004, 18080 Granada, Spain. (fjpi@iaa.es)
}

\authoraddr{A. Luque,
Instituto de Astrof\'isica de Andaluc\'ia (IAA),
CSIC, PO Box 3004, 18080 Granada, Spain. (aluque@iaa.es)
}

\authoraddr{F. J. Gordillo-V\'azquez,
Instituto de Astrof\'isica de Andaluc\'ia (IAA),
CSIC, PO Box 3004, 18080 Granada, Spain. (vazquez@iaa.es)
}

\begin{abstract}

The spatial non-uniformity of the electric field in air discharges, such as streamers, can influence the accuracy of spectroscopic diagnostic methods and hence the estimation of the peak electric field. In this work, we use a self-consistent streamer discharge model to investigate the spatial non-uniformity in streamer heads and streamer glows. We focus our analysis on air discharges at atmospheric pressure and at the low pressure of the mesosphere. This approach is useful to investigate the spatial non-uniformity of laboratory discharges as well as sprite streamers and blue jet streamers, two types of Transient Luminous Event (TLE) taking place above thunderclouds. This characterization of the spatial non-uniformity of the electric field in air discharges allows us to develop two different spectroscopic diagnostic methods to estimate the peak electric field in cold plasmas. The commonly employed method to derive the peak electric field in streamer heads underestimates the electric field by about 40-50~\% as a consequence of the high spatial non-uniformity of the electric field. Our diagnostic methods reduce this underestimation to about 10-20\%. However, our methods are less accurate than previous methods for streamer glows, where the electric field is uniformly distributed in space. Finally, we apply our diagnostic methods to the measured optical signals in the Second Positive System of $N_2$ and the First Negative System of $N_2^+$ of sprites recorded by Armstrong et al. (1998) during the SPRITE's 95 and 96 campaigns.

\end{abstract}

\begin{article}

\section{Introduction}
\label{sec:intro}
 Non-equilibrium (or non-thermal) air discharges are due to the application of an electric field, which provides energy to the electrons and maintains the ionization of air molecules. In non-equilibrium discharges, the electron temperature exceeds the background temperature. The discharge parameters that determine the type of discharge are their spatial and temporal scales, the production of  electron avalanches and the plasma and air temperature. We refer to \cite{bruggeman2017/PSST} for a detailed description of the different types of non-thermal air discharges.

Non-equilibrium air discharges have numerous industrial applications \citep{simek2014optical} and are closely related to atmospheric electricity phenomena, such as lightning and Transient Luminous Events (TLEs) \citep{Franz1990/Sci, Pasko2012/SSR}. TLEs are upper atmospheric discharges related to lightning. Sprites and blue jets are TLEs that occur above thunderstorms that cover altitudes ranging between 20~km and 85~km \citep{Wescott1995/GeoRL, Wescott1996/GeoRL/1, Pasko1996/GeoRL, Wescott1998/JASTP/1, Stenbaek-Nielsen2000/GeoRL/corr, Wescott2001/JGR, Gordillo-Vazquez2009/PSST, Gordillo-Vazquez2018/JGR}. The lower part of sprites and the upper part of blue jets are formed by hundreds of streamer discharges \citep{Parra2014/JGR,Kuo2015/JGR, luque2016sprite} and emit light predominantly in some band systems of molecular nitrogen. Sprites are one of the largest non-thermal air discharges in nature. For a more extensive review of TLEs and sprites, we refer to \cite{Pasko2012/SSR}.

\cite{Gallimberti1974/JPhD} and \cite{Goldman1978/gael.conf}  (1978, p. 243) investigated the optical spectra of air corona discharges at atmospheric pressure, noting that the spectra were dominated by N$_2$ emissions. Later investigations (see \cite{stritzke1977spatial, kondo1980highly} and references collected by \cite{simek2014optical}) confirmed that non-equilibrium air discharges emit light predominantly in the first and second positive band systems of the molecular neutral nitrogen (1PS~N$_2$ and the 2PS~N$_2$, or simply FPS and SPS), the first negative band system of the molecular nitrogen ion (N$_2$$^+$-1NS or simply FNS), the Meinel band system of the molecular nitrogen ion (Meinel N$_2$$^+$) and the Lyman-Birge-Hopfield (LBH) band system of the molecular neutral nitrogen. In this work, we will refer to FNS, FPS and SPS for the sum of optical emissions between all the vibrational states. The emissions from the vibrational state $v^{\prime}=v_i$ to all $v^{\prime\prime}$ will be labeled as FNS$_{(v_i, v^{\prime\prime})}$, FPS$_{(v_i, v^{\prime\prime})}$ and SPS$_{(v_i, v^{\prime\prime})}$.  Finally, the emissions from the vibrational state $v^{\prime}=v_i$ to $v^{\prime\prime}=v_j$ will be labeled as FNS$_{(v_i, v_j)}$, FPS$_{(v_i, v_j)}$ and SPS$_{(v_i, v_j)}$.

The electronic excitation thresholds for the production of N$_2$$(B^{3}\Pi_g, v=0$), N$_2$$(C^{3}\Pi_u, v=0)$, N$_2$$^+$$(B^2\Sigma_u^+, v=0)$ are respectively 7.35~eV, 11.03~eV and 18.80~eV \citep{Phelps1985/PhRvA}. Therefore, the population of these emitting molecules in air discharges depends on the electric field that provides energy to the electrons. Excited molecules N$_2$$(B^{3}\Pi_g, v=0$), N$_2$$(C^{3}\Pi_u, v=0)$ and N$_2$$^+$$(B^2\Sigma_u^+, v=0)$ can respectively emit photons in the optical FPS, SPS and FNS. \cite{creyghton1994pulsed} proposed the use of the intensity ratio of FNS to SPS to estimate the peak electric field that produces these molecular excitation in streamers. Other authors have also used this intensity ratio to estimate the peak electric field in air discharges \citep{kozlov2001spatio, Morrill2002/GeoRL, kim2003measurements, paris2004measurement, Paris2005/JPhD, Kuo2005/GRL, shcherbakov2007subnanosecond, Kuo2009/JGR, Kuo2013/JGRA, Adachi2006/GeoRL, Liu2006/GeoRL, Pasko2010/JGRA, Celestin2010/GeoRL, Bonaventura2011/PSST, hoder2012high, Holder2016/PSST}. The ratio of FPS to SPS has also been proposed to calculate the peak electric field in air discharges \citep{simek2014optical}. \cite{ihaddadene2017determination} proposed a spectroscopic diagnostic method to derive the altitude of sprites streamers based on the altitude dependence of the quenching rate of different electronic excited states of N$_2$.

\cite{creyghton1994pulsed} and \cite{Naidis2009/PhRvE} noted that the peak electric field obtained from the ratio of FNS to SPS in streamer discharges is distorted by the spatial non-uniformity of the streamer head. \cite{Celestin2010/GeoRL} used a self-consistent streamer model to calculate the synthetic optical emissions of positive and negative streamers. They compared the peak electric field calculated by the model with the peak electric field estimated from the synthetic optical emissions. According to \cite{Celestin2010/GeoRL}, the peak electric field obtained from the ratio of FNS to SPS in streamer discharges must be multiplied by a factor $\Gamma_E$ ranging between 1.4 and 1.5.

Apart from the spatial non-uniformity of the streamer discharge, the uncertainty of the reaction rates involved in the discharge can lead to a significant error in the estimated electric field \citep{creyghton1994pulsed, kozlov2001spatio, paris2004measurement, Holder2016/PSST}. Recently, \cite{obrusnik2018electric} and \cite{bilek2018electric} performed a sensitivity analysis to determine the effect of the reaction rate uncertainties in the obtained peak electric field at different pressures using the ratio of FNS to SPS. According to their results, the processes that significantly influence the error in the estimated peak electric field from optical emissions are the excitation by electron impact, the radiative de-excitation and the electronic quenching by air of electronically excited states of N$_2$ and N$_2$$^+$, especially at atmospheric pressure. \cite{simek2014optical} reported an additional error in the estimated peak electric field from the ratio of FPS to SPS as a consequence of the electric field dependence of the Vibrational Distribution Function of N$_2$$(B^{3}\Pi_g, v^{\prime})$ at relatively low electric field values ($\sim$150-200~Td). Finally, \cite{simek2014optical} and \cite{Perezinvernon2018b/JGR} demonstrated that the ratio of FPS to SPS is highly inaccurate for reduced electric field values above $\sim$200~Td, as this ratio is almost electric field independent for higher electric field values.

In this work, we develop two methods to reduce the uncertainty in the estimated peak electric field caused by the spatial non-uniformity of the discharge using the ratio of FNS to SPS and FPS to SPS. Firstly, we use a streamer model to calculate the synthetic optical emissions of a laboratory streamer head, a sprite streamer head and a sprite streamer glow. By sprite streamer glow we refer to the column-like luminous structure that appears in the sprite streamer wake after the streamer head passage  \citep{Stenbaek-Nielsen2008/JPhD,Gordillo-Vazquez2010/GeoRL,Luque2010/GeoRL,Liu2010/GeoRL,luque2016sprite}. Secondly, we use the model to analyze the non-uniformity of the discharges. Finally, we use the synthetic optical emissions and the spatial non-uniformity of the discharges to make more accurate the commonly used method to calculate the peak electric field in the plasma.

The organization of this paper is as follows: Section~\ref{sec:streamermodel} briefly describes the streamer model used to generate the synthetic optical emissions of streamer heads and a glow. Sections~\ref{sect:signal} and \ref{sec:comparison} are devoted to the improved diagnostic methods for
non-uniform air discharges. Section~\ref{sec:signal_armostrong} highlights the applicability of the methods in the analysis of optical emissions from TLEs reported by \cite{Armstrong1998/JASTP}. The conclusions are finally presented in section~\ref{sec:conclusions}.

\section{Streamer model}
\label{sec:streamermodel}
Our model is 2D cylindrically symmetric and the dynamics of all charged species is described
by diffusion-drift-reaction equations for electrons and ions coupled with Poisson's equation as follows,

\begin{subequations}
\label{eq:streamer_model}
\begin{equation}
\frac{\partial n_{e}}{\partial t} = \nabla\cdot\left(n_{e}\mathbf{\mu_e\mathbf{E}} + D_{e}\nabla n_{e}\right) + C_{e} +  S_{ph},\label{eq:number_density}
\end{equation}

\begin{equation}
\frac{\partial n_{i}}{\partial t} = C_{i} + S_{ph},\label{eq:number_density_ions}
\end{equation}

\begin{equation}
-\nabla \cdot \mathbf{E} = \nabla^{2}\phi = -\frac{\rho}{\epsilon_0},\label{eq:poisson_equation}
\end{equation}
\end{subequations}

where $n_{e,i}$ is the number density for electrons and ions respectively, $\mu_e$ is the electron mobility and $D_{e}$ is the diffusion coefficient. In the present model we consider ions motionless over the short timescales  that we study and therefore we neglect mobility and diffusion coefficients of ions. The term $C_{s=e,i}$  is the net production of species $s$ due to chemical processes, and $S_{ph}$ is the photoionization term that we calculate following the procedure described by \cite{Luque2007/ApPhL}. Photoionization acts only on the densities of $e$ and $O_2^+$.  As for Poisson's equation, $E$ is the electric field, $\phi$ is the electrostatic potential, $\rho$ is the density of charges and $\epsilon_0$ is the permittivity of vacuum. In this work we use the local field approximation and therefore, transport coefficients are derived from the electron energy distribution function (EEDF) that depends only on the local electric field.

The streamers develop in a $N_2$:$O_2$-mixture (79:21)  and the basic kinetic scheme  accounts for impact ionization and attachment/detachment, as described in the supplementary material by \cite{Luque2017/PSST}, but excluding the water chemistry. Some of the collisions that the electrons undergo excite molecules electronically and vibrationally. These excited molecules either decay emitting a photon of a given frequency or are collisionally quenched, i.e. decay to a fundamental level through collisions with $N_2$ and $O_2$. In order to account for these emissions we include electronic and vibrational excitations and de-excitations as well as radiative decay and quenching. Table~\ref{table:op_reac} summarizes the most important processes that influence the optical emissions.

Finite Volume Methods are suitable to solve the set of equations (\ref{eq:streamer_model}). To solve these equations we have used CLAWPACK/PETCLAW \citep{LeVeque2002/book,Alghamdi2011/PSP,clawpack}. PETCLAW is built upon PETSc \citep{PETSc2016/web,PETSc2016/user} and allows us to split the simulation domain into different subdomains (problems) that can be solved in parallel. Poisson's equation is solved using the Generalized Minimal Residual method and the geometric algebraic multigrid  preconditioner, both from the PETSc numerical library.

\small
\begin{table}
\caption{Most important processes for the optical emissions.} \label{table:op_reac}
\scalebox{0.58}{
\begin{tabular}{l*{3}{c}r}
\hline
       Chemical reaction       & Rate & Reference  \\
\hline
e + N$_2$$(X^1\Sigma_g^+, v=0)$ $\rightarrow$ N$_2$$^+$$(B^2\Sigma_u^+, v^{\prime}=0)$ + 2e & $k_{B^2\Sigma_u^+, 0}$ = $f\left(\frac{E}{N}\right)$ & (\cite{Hagelaar2005/PSST,Phelps1985/PhRvA}) \\
e + N$_2$$(X^1\Sigma_g^+, v=0)$ $\rightarrow$ N$_2$$^+$$(B^2\Sigma_u^+, v^{\prime}=1)$ + 2e & $k_{B^2\Sigma_u^+, 1}$ = $f\left(\frac{E}{N}\right)$ & (\cite{Hagelaar2005/PSST,Phelps1985/PhRvA})  \\
e + N$_2$$(X^1\Sigma_g^+, v=0)$ $\rightarrow$ N$_2$$(B^{3}\Pi_g, v^{\prime})$ + e & $k_{B^{3}\Pi_g}$ = $f\left(\frac{E}{N}\right)$ & (\cite{Hagelaar2005/PSST,Phelps1985/PhRvA})  \\
e + N$_2$$(X^1\Sigma_g^+, v=0)$ $\rightarrow$ N$_2$$(C^{3}\Pi_u, v^{\prime})$ + e & $k_{C^{3}\Pi_u}$ = $f\left(\frac{E}{N}\right)$ & (\cite{Hagelaar2005/PSST,Phelps1985/PhRvA})  \\

N$_2$$^+$$(B^2\Sigma_u^+, v^{\prime}=0)$  $\rightarrow$ N$_2$$^+$$(X^2\Sigma_g^+, v^{\prime\prime}=0)$ + $h\nu$(FNS$_{(0,v^{\prime\prime})}$) & $A_{B^2\Sigma_u^+ (0,v^{\prime\prime})}$ =  1.14 $\times$ 10$^7$ s$^{-1}$ & (\cite{Gilmore1992/JPCRD})  \\
N$_2$$^+$$(B^2\Sigma_u^+, v^{\prime}=1)$  $\rightarrow$ N$_2$$^+$$(X^2\Sigma_g^+, v^{\prime\prime}=0)$ + $h\nu$(FNS$_{(1,v^{\prime\prime})}$) & $A_{B^2\Sigma_u^+ (1,v^{\prime\prime})}$ = 3.71 $\times$ 10$^6$ s$^{-1}$ & (\cite{Gilmore1992/JPCRD})  \\
N$_2$$(B^{3}\Pi_g, v^{\prime})$  $\rightarrow$ N$_2$$(X^1\Sigma_g^+, v^{\prime\prime})$  + $h\nu$(FPS) & $A_{B^{3}\Pi_g}$ = 1.34 $\times$ 10$^5$ s$^{-1}$ & (\cite{Capitelli2000/book})  \\
N$_2$$(C^{3}\Pi_u, v^{\prime})$  $\rightarrow$ N$_2$$(B^{3}\Pi_g, v^{\prime\prime})$  + $h\nu$(SPS) & $A_{C^{3}\Pi_u}$ = 2.47 $\times$ 10$^7$ s$^{-1}$ & (\cite{Capitelli2000/book})  \\

N$_2$$^+$$(B^2\Sigma_u^+, v=0)$ + M  $\rightarrow$ N$_2$$^+$$(X^2\Sigma_g^+, v=0)$ + M & $Q_{B^2\Sigma_u^+, N_2}$ = 8.84 $\times$ 10$^{-10}$ cm$^{3}$s$^{-1}$  $Q_{B^2\Sigma_u^+, O_2}$ = 10.45 $\times$ 10$^{-10}$ cm$^{3}$s$^{-1}$  & (\cite{Dilecce2010/JPhD/c})  \\
N$_2$$^+$$(B^2\Sigma_u^+, v=1)$ + M  $\rightarrow$ N$_2$$^+$$(X^2\Sigma_g^+, v=0)$ + M & $Q_{B^2\Sigma_u^+, N_2}$ = \SI{16e-10}{cm^{-3}s^{-1}} $Q_{B^2\Sigma_u^+, O_2}$ = \SI{10.45e-10}{cm^{-3}s^{-1}} & (\cite{Jolly1983/PSST})  \\
N$_2$$(B^{3}\Pi_g, v)$ + M  $\rightarrow$ Deactivated products & $Q_{B^{3}\Pi_g, N_2}$ = 2 $\times$ 10$^{-12}$ cm$^{3}$s$^{-1}$ $Q_{B^{3}\Pi_g, O_2}$ = 3 $\times$ 10$^{-10}$ cm$^{3}$s$^{-1}$ & (\cite{Capitelli2000/book})  \\
N$_2$$(C^{3}\Pi_u, v)$ + M $\rightarrow$ Deactivated products & $Q_{C^{3}\Pi_u, N_2}$ = 10$^{-11}$ cm$^{3}$s$^{-1}$ $Q_{C^{3}\Pi_u, O_2} $ = 3 $\times$ 10$^{-10}$ cm$^{3}$s$^{-1}$ & (\cite{Capitelli2000/book})  \\
\end{tabular}
}
\end {table}
\normalsize

\subsection{Sprite streamer}
Sprites are high-altitude discharges made of many streamers that propagate through a varying air density. In our model,
the air density follows a
decaying exponential profile with an $e$-folding length of 7.2 km. We also set a background electron density following the Wait-Spies profile:

\begin{equation}
n_{e,bg} =\left( \SI{e4}{m^{-3}}\right) \times e^{-\left(z-\SI{60}{km}\right)/\SI{2.86}{km}} \label{Wait-Spies}
\end{equation}

\begin{figure}
\centering
\includegraphics[width=18cm]{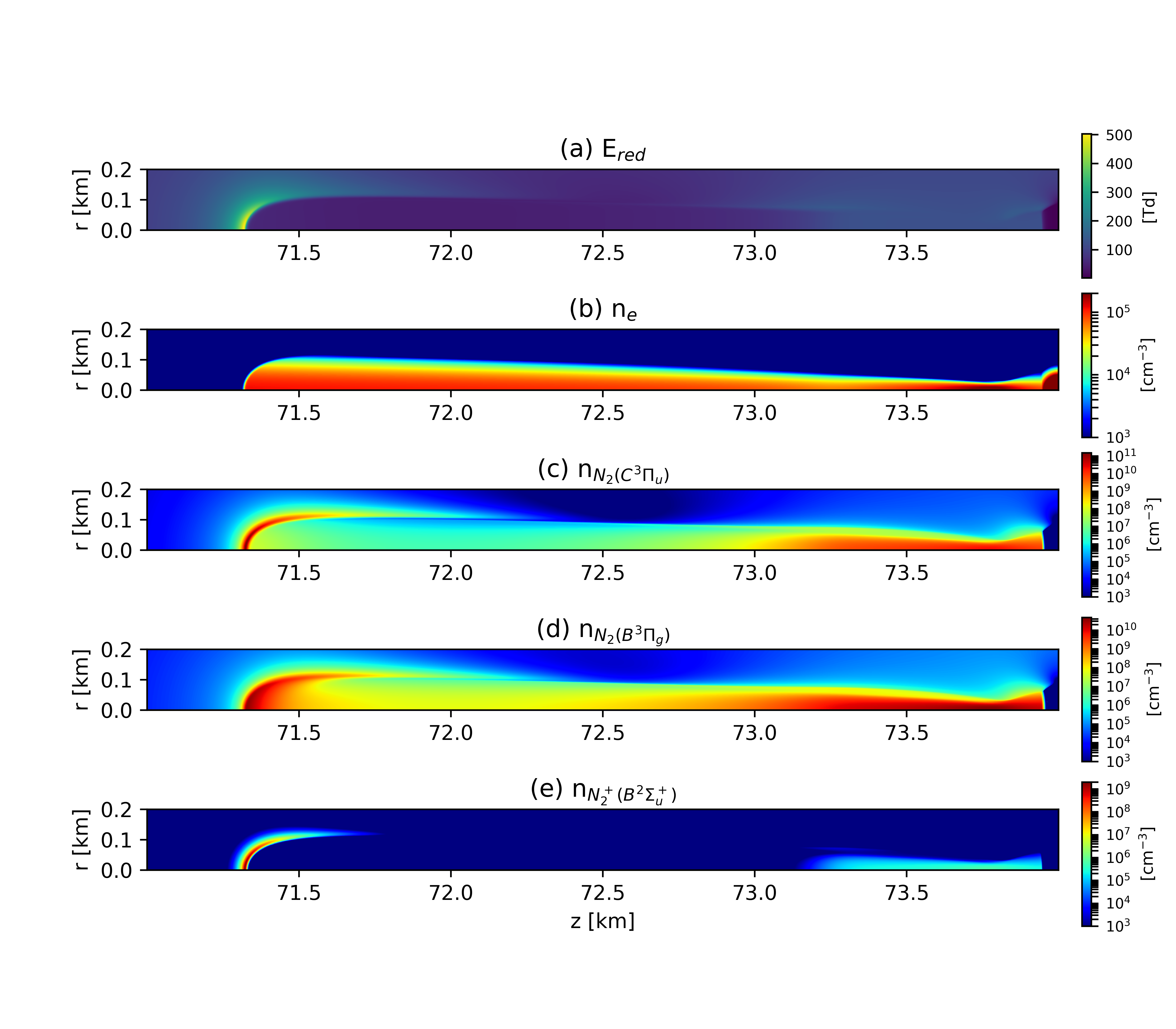}
\caption{ Cross-sectional view of (a) reduced electric field, (b) electron density and (c-e) density of emitting molecules for a downward propagating positive sprite streamer and a glow for a background electric field of 100~V/m at time 0.9~ms.}
\label{fig:plot_sprite}
\end{figure}

In order to start the streamer we set a gaussian seed with an $e$-folding radius of $\SI{20}{m}$  and a peak
density of $\SI{5e12}{m^{-3}}$. This initial electron density is neutralized by an identical density of positive ions.
In order to solve Poisson's equation we impose Dirichlet boundary conditions at $z=z_{min},z_{max}$ and free boundary conditions at $r=r_{max}$
according to the method described by \cite{Malagon-Romero2018/CoPhC}.
These free boundary conditions are consistent with the density charge inside the domain and with a potential decaying far away from the source.
We have simulated positive and negative streamers propagating in background electric fields of 100 V/m and 120 V/m, which correspond to 120 Td at 74.23 km and 72.91 km respectively. The simulated domain extends from 71 to 75 km in the vertical direction and the grid resolution is 1 m.
We have calculated the optical emissions from the streamer head in a moving cylindrical box of radius $\SI{150}{m}$ and height $\SI{150}{m}.$ We have also calculated the optical emissions from the glow in a cylindrical box of radius $\SI{30}{m}$ and height $\SI{700}{m}$. Figure~\ref{fig:plot_sprite} shows a simulated positive streamer propagating downward for a background electric field of 100~V/m and a glow emerging at $z$ = 73.5~km.

\subsection{Laboratory streamer}
The ground-level streamer discharge develops in a needle-plane configuration.
Our initial condition consists in a needle with an small ionization patch slightly off the needle tip.
The needle is simulated by a narrow elongated volume with a high ionization.
The initial electron density is thus the sum of a uniform background $n_e^\text{bg}=\SI{e9}{m^{-3}}$ plus
\begin{equation}
  n_e^{\text{needle}} = n_{e0} \exp\left(-\frac{\max\left(z - z_n ,0\right)^2}{\sigma_n^2} - \frac{r^2}{\sigma_n^2}\right),
  \label{initial_leader}
\end{equation}
and
\begin{equation}
  n_e^{\text{seed}} = n_{e0}  \exp\left(-\frac{\left(z - z_S\right)^2}{\sigma_S^2} - \frac{r^2}{\sigma_S^2}\right),
  \label{initial_streamer}
\end{equation}
where $z_n=\SI{5.4}{mm}$ is the tip location,
$z_S=\SI{7.5}{mm}$ is the center of the seed,  $\sigma_n = \SI{0.9}{mm}$
 and $\sigma_S = \SI{0.45}{mm}$ are the $e$-folding radii
 and the electron density peaks at $n_{e0}= \SI{e21}{m^{-3}}$.
The initial electron density is neutralized by an identical density of positive ions. Boundary conditions are the same as in the sprite streamer simulation. We have simulated positive and negative streamers with background electric fields $E_{bg}=\lbrace12.5,15,20,25\rbrace$ kV/cm and $E_{bg}=\lbrace15,20,25\rbrace$ kV/cm respectively. The full domain size is $\SI{3}{cm}\times\SI{1}{cm}$ and the grid resolution is $\SI{5}{\micro m}$. We have calculated the optical emissions in a moving cylindrical box of radius $\SI{3.2}{mm}$ and a vertical extension between 0 and $\SI{3.2}{mm}$ containing the streamer head.

\section{Spatial non-uniformity of the electric field in spectroscopic diagnostics}
\label{sect:signal}

In non-thermal air discharges optical emissions are mainly determined by the concentration of electronically excited nitrogen molecules. The plasma involved in such electrical discharges is far from thermal equilibrium, ensuring that electron-impact processes driven by the electric field are responsible for the excitation of the emitting molecules. Radiative de-excitation processes together with other chemical reactions, such as electronic quenching by air, contribute to the de-excitation of the excited molecules. The total number of emitted photons in non-equilibrium gas discharges depends on the reduced electric field and the competition between radiative de-excitation and other de-excitation processes.

In this section we develop and compare two methods to estimate the peak electric field in a non-equilibrium plasma by considering the effect of the non-uniformity of the electric field. These methods pursue an estimation of the peak electric field from the ratio of optical emissions of different nitrogen band systems based on the spatial non-uniformity of the electric field (see \cite{simek2014optical} and references therein). In principle these methods can be generalized to other gases than air if appropriate emission lines are identified.

\subsection{Peak electric field from the relation between electron and electric field spatial distributions}
\label{sec:method}


The density of the emitting species in a non-equilibrium plasma $N_s(t)$ can be estimated from the decay constant $A^{\prime}$ of the transitions that produce photons in a considered wavelength range and the observed intensity $I(t)$  as

\begin{equation}
N_s(t) = \frac{I(t)}{A^{\prime}}. \label{densities}
\end{equation}

The temporal production rate $S(t)$ due to electron impact can be derived from the continuity equation of the emitting species as

\begin{equation}
S(t) = \frac{dN_s(t)}{dt} + A  N_s(t) + Q N_s(t) N - C N^{\prime}(t) + R(t), \label{productionc}
\end{equation}

where $A$ is the total radiative decay constant of the emitting species, $Q$ represents all the quenching rate constants by air molecules of density $N$. $N^{\prime}(t)$ accounts for the density of all the upper species that populate the species by radiative cascade with radiative decay constants $C$. Finally, the term $R(t)$ includes the remaining loss processes, such as intersystem processes or vibrational redistribution. \cite{obrusnik2018electric} and \cite{bilek2018electric} demonstrated that the most important processes that influence the optical emissions are the excitation of emitting molecules by electron impact, radiative de-excitations and electronic quenching. Therefore, we can neglect the effect of other processes in the derivation of the peak reduced electric field and approximate equation~(\ref{productionc}) as

\begin{equation}
S(t) \simeq \frac{dN_s(t)}{dt} + A  N_s(t) + Q N_s(t) N. \label{production}
\end{equation}

From equations~(\ref{densities}) and~(\ref{production}) we can obtain the production ratios of two different species (1 and 2) at a fixed time $t$, given by $S_{12} = \frac{S_1}{S_2}$ as a first approximation, without considering any spatial non-uniformity of the electric field. The magnitude $S_{12}$ and the electron-impact production ratio of species 1 and 2 given by $\nu_{12} = \frac{\nu_1(E/N)}{\nu_2(E/N)}$, allow us to estimate the reduced electric field that satisfies the equation

\begin{equation}
\frac{S_1}{S_2} \simeq \frac{\nu_1(E/N)}{\nu_2(E/N)} . \label{equality}
\end{equation}

We get the values of $\nu_i(E/N)$ for all the considered species using BOLSIG+ for air \citep{Hagelaar2005/PSST}. This common approximation is useful to estimate the electric field value at the point where the rate of excitation is maximal ($\varepsilon^{\prime}$). However,  $\varepsilon^{\prime}$ is only equal to the peak electric field value in the discharge as long as the electric field is uniform.
Optical emissions from non-thermal air discharges are generally produced by inhomogeneous electric fields and streamers heads are a clear example \citep{Naidis2009/PhRvE, Celestin2010/GeoRL}. \cite{Perezinvernon2018b/JGR} investigated the spatial non-uniformity of the electric field and its effect on the optical emissions of halos and elves, two kinds of diffuse TLEs. They defined the function $H(\varepsilon)$ as the number of electrons under the influence of a reduced electric field (defined as $E_{red}$ = $\frac{E}{N}$) larger than $\varepsilon$ and weighted by the air density $N$

\begin{equation}
H(\varepsilon) = \int d^{3}\mathbf{r}~N(\mathbf{r})~n_e(\mathbf{r})~\theta\left( E_{red}(\mathbf{r}) - \varepsilon \right), \label{nen}
\end{equation}

where $n_e(\mathbf{r})$ and $E_{red}(\mathbf{r})$ are, respectively, the electron density and the reduced electric field spatial distributions and the integral extends over all the volume of the discharge. The symbol $\theta$ corresponds to the step function, being 1 if $E_{red} > \varepsilon$ or 0 in any other case. The total excitation rate of species $i$ by electron impact in the spatial region occupied by the discharge can be written as a function of $H$ as

\begin{equation}
\nu_i = \int_{E_{red, min}}^{E_{red, max}} d \varepsilon \left| \frac{dH}{d \varepsilon } \right| k_i\left( \varepsilon \right), \label{productioni_1}
\end{equation}

where $E_{red, max}$ and $E_{red, min}$ are, respectively, the maximum and the minimum reduced electric field in the region where the optical emissions are produced and $k_i\left( \varepsilon \right)$ is the reaction rate coefficient for electron-impact excitation of species $i$.



The function defined by equation~(\ref{nen}) contains information about the spatial non-uniformity of the discharge. \cite{Perezinvernon2018b/JGR} found that the function $H(\varepsilon)$ can be approximated as a linear function for halos and elves, as the electron density is not significantly affected by the electric field in those events. However, high values of the electric field in streamer heads produce an enhancement of some orders of magnitude in the background electron density. We have used the streamer model described in section~\ref{sec:streamermodel} to find a general approximation to this function so we can use it in streamer and glow discharges. In particular, our approximation must fit the modeled curve in the electric field range where the maximum excitation of emitting molecules occurs. Examination of equation~(\ref{productioni_1}) indicates that the maximum excitation is produced at the electric field value where the product between the derivative $\frac{dH}{d \varepsilon }$ and the electron-impact excitation rate coefficient reaches its maximum. Then, the spatial non-uniformity of the electric field can influence the value of the electric field that produces the maximum excitation.

\begin{figure}
\centering
\includegraphics[width=8cm]{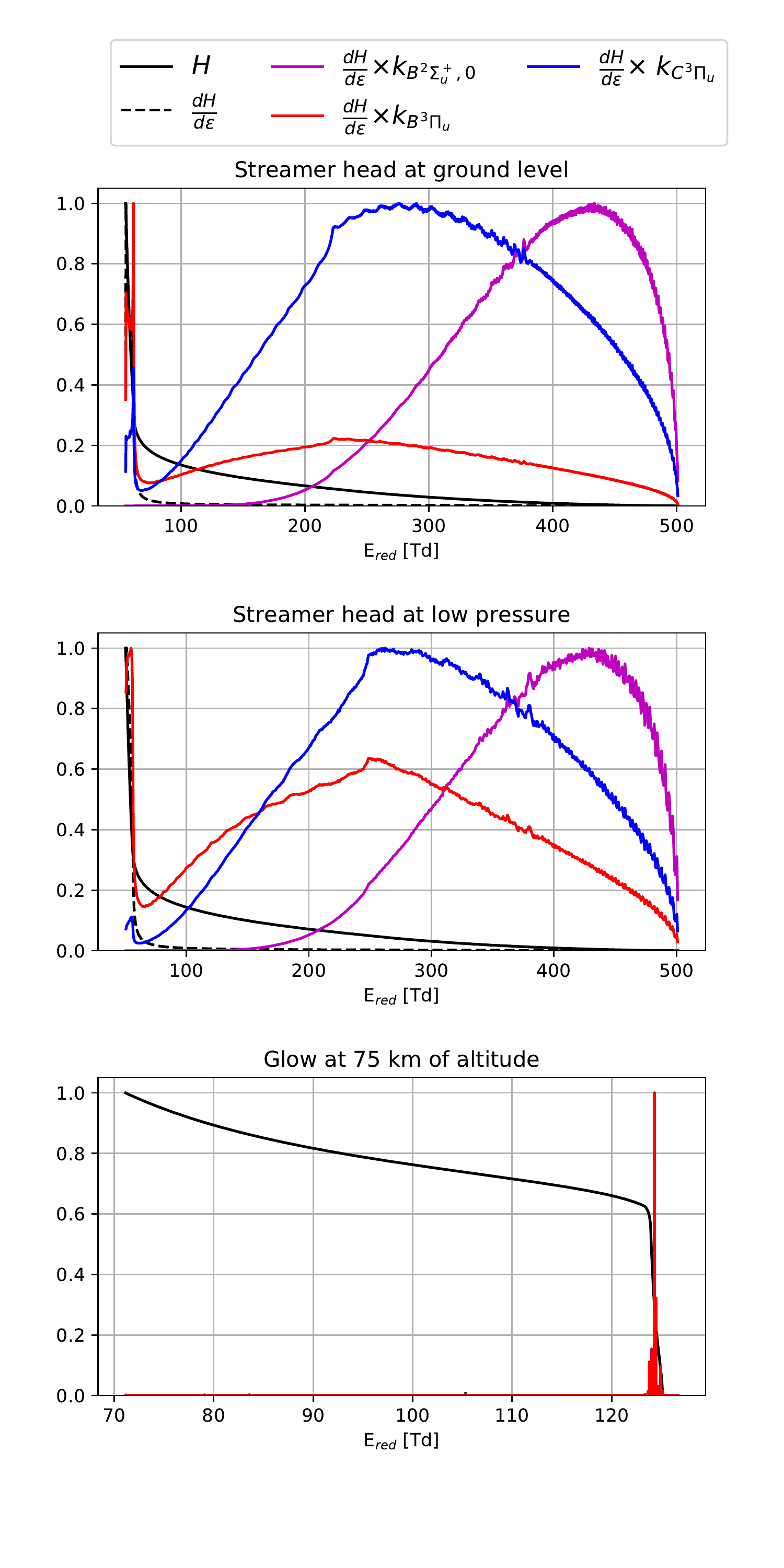}
\caption{Function $H$ (black solid lines), $\frac{dH}{d \varepsilon }$ (dashed black lines) and product between $\frac{dH}{d \varepsilon }$ and the electron-impact reaction rate coefficient $k_i$ indicating the electric field range where excitation of emitting molecules is important (color solid lines). We plot the results for a positive streamer at atmospheric pressure with a background electric field of 20~kV~cm$^{-1}$ and 65~ns after its onset (first panel), a positive streamer at low pressure with a background electric field of 100~V~m$^{-1}$ and 0.9~ms after its onset (second panel) and its glow 0.9~ms after its onset (third panel). All curves are normalized to their maximum value.}
\label{fig:HE}
\end{figure}

Solid black lines in figure~\ref{fig:HE} show the normalized function $H$ (equation~(\ref{nen})) obtained for two streamer heads simulated at different pressures and for a low pressure streamer glow. Dashed black lines in figure~\ref{fig:HE} correspond to the normalized derivative $\frac{dH}{d\varepsilon}$. Finally, color solid lines in figure~\ref{fig:HE} show the normalized product between $\frac{dH}{d\varepsilon}$ and different electron-impact excitation rate coefficients. The approximation to $H$ must fit the solid black lines in figure~\ref{fig:HE} in the electric field range where the product between $\frac{dH}{d\varepsilon}$ and the electron-impact excitation rate coefficient is greater than zero. It can be clearly seen in figure~\ref{fig:HE} that most emissions produced in the glow are located in the region where the electric field reaches its maximum value. Therefore, the method neglecting the spatial non-uniformity to estimate the peak electric field is accurate enough to study glows. However, the situation is different  in streamer heads, where the maximum excitation is not produced in the region where the electric field reaches its maximum. Therefore, we need to develop a method accounting for the spatial non-uniformity of the electric field distribution.

By definition, the function $H$ is constant between 0~Td and the minimum field that influences electrons, $E_{red, min}$. Figure~\ref{fig:HE} shows that for streamer heads, $H$ decreases between $E_{red, min}$ and $E_{red, max}$, while $H(E_{red, max})$ = 0 by definition. Regarding halos and elves \citep{Perezinvernon2018b/JGR}, $H$ could be approximated as a linear function. However, a higher-order approximation is convenient in the case of streamer heads. We have examined the electric field-dependence of $H$ concluding that it can be approximated in general as

\begin{equation}
H\left(\varepsilon\right) \simeq \alpha \left( E_{red, max}- \varepsilon \right)^{\beta}, \label{NEN_line}
\end{equation}

where $\alpha$ and $\beta$ are constants.

\begin{figure}
\centering
\includegraphics[width=8cm]{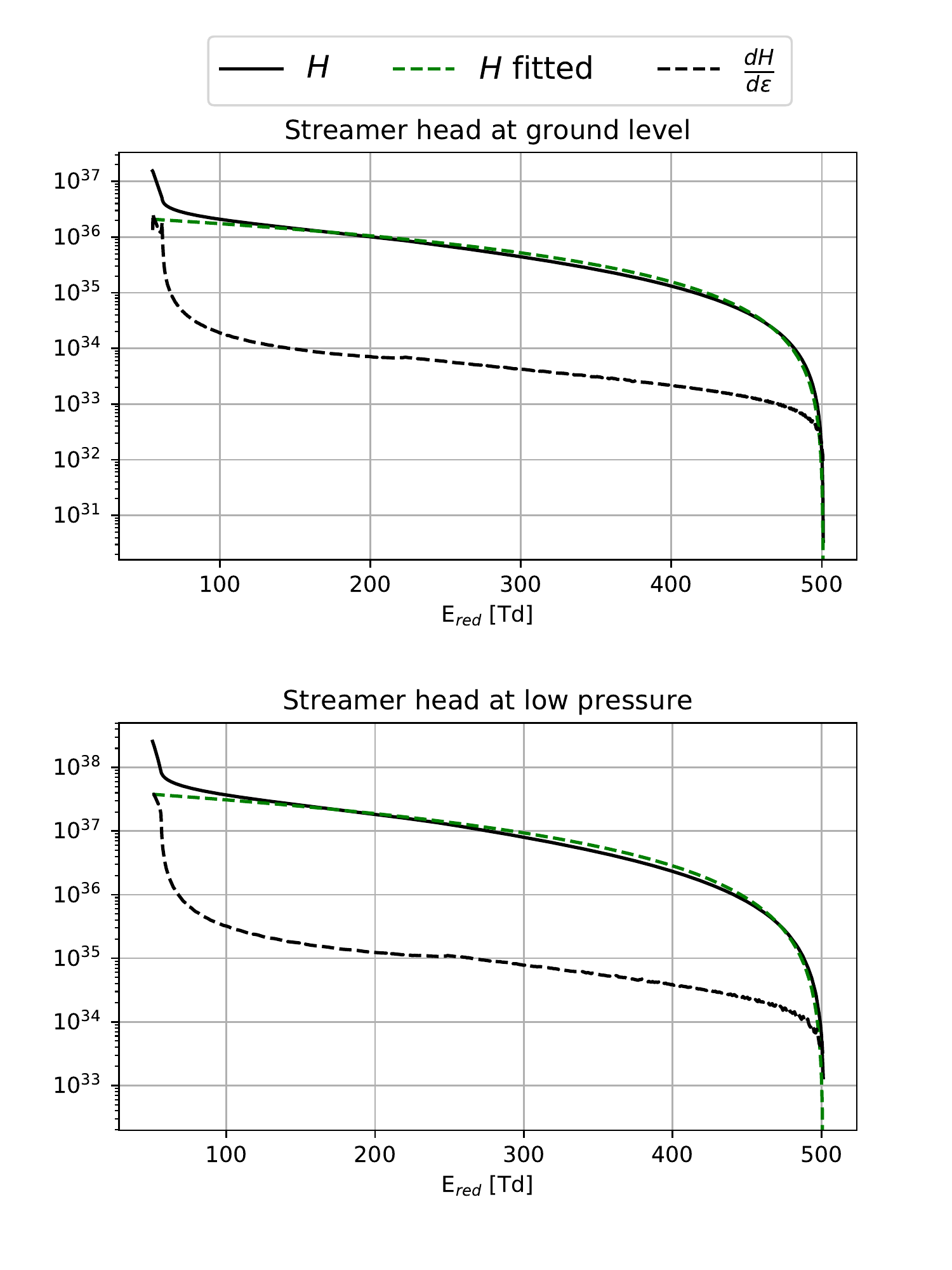}
\caption{Reduced electric field dependence of the function $H$ (black solid lines), the fitting of  $H$ using equation~(\ref{NEN_line}) (dashed green lines) and $\frac{dH}{d \varepsilon }$ (dashed black lines). We plot the results for a positive streamer at atmospheric pressure with a background electric field of 20~kV~cm$^{-1}$ and 65~ns after its onset (first panel) and for a positive streamer at low pressure with a background electric field of 100~V~m$^{-1}$ and 0.9~ms after its onset (second panel).}
\label{fig:HE_fit}
\end{figure}

We have performed 7 negative and positive laboratory-like streamer simulations with different background electric fields (12.5~kV~cm$^{-1}$, $\pm$15~kV~cm$^{-1}$, $\pm$20~kV~cm$^{-1}$ and $\pm$25~kV~cm$^{-1}$) and 4 negative and positive sprite-like streamer simulations (71 to 75~km) with different background electric fields ($\pm$100~V~m$^{-1}$ and $\pm$120~V~m$^{-1}$). The value of the obtained functions $H$ ranges between zero and several orders of magnitude in all cases. Therefore, we have used a logarithmic least square fitting of equation~(\ref{NEN_line}) in order to minimize the error of the coefficients $\alpha$ and $\beta$. The fitting of equation~(\ref{NEN_line}) has been performed in a equispaced grid of electric field values to ensure that the minimization of the error weighting does not depend on the distribution of points of the functions $H$ calculated by the streamer model. We plot the function $H$ together with the obtained fitting and the derivative $\frac{dH}{d\varepsilon}$ for laboratory and sprite streamer heads in figure~\ref{fig:HE_fit}. We have obtained that the $\beta$ exponent is between 1.8 and 2.0 with a mean squared error of about 6$\times$10$^{-3}$ for all the laboratory streamers and between 1.95 and 2.09 with a mean squared error of about 10$^{-2}$ for all the sprite streamers. Thus, these values do not significantly depend on the streamer polarity, background electric field or pressure. Consequently, we take the average value $\beta$ = 1.96. As the $\beta$ exponent is always close to 2, we will refer to this diagnostic method as ``Quadratic Method". The obtained values of $\alpha$ are about 5$\times$10$^{31}$ with a mean squared error of about 1.5$\times$10$^{30}$ for laboratory streamers and about 10$^{33}$ with a mean squared error of about 5$\times$10$^{31}$ for sprite streamers.

Let us now deduce the expression for the production of emitting species by electron impact considering this approximation to $H$. Applying the derivative to equation (\ref{NEN_line}), equation~(\ref{productioni_1}) can be written as

\begin{equation}
\nu_i = \beta \, \alpha \int_{E_{red, min}}^{E_{red, max}} d \varepsilon  \left( E_{red, max} - \varepsilon \right)^{\beta-1} k_i\left( \varepsilon \right). \label{productioni}
\end{equation}

Now, following equations (\ref{equality}) and (\ref{productioni}), we write the  production rate ratio of two species by electron impact derived from the recorded optical intensity $\left(S_{12} = \frac{S_1}{S_2}\right)$ as

\begin{equation}
S_{12} = \frac{S_1}{S_2} \simeq \frac{ \int_{E_{red, min}}^{E_{red, max}} d \varepsilon  \left( E_{red, max} - \varepsilon \right)^{\beta-1} k_1\left( \varepsilon \right)} {\int_{E_{red, min}}^{E_{red, max}} d \varepsilon  \left( E_{red, max} - \varepsilon \right)^{\beta-1} k_2\left( \varepsilon \right)}. \label{production_ratio}
\end{equation}

In order to derive $E_{red,max}$ we need to know the minimum electric field in the region where the optical emissions are produced. In general, $E_{red,min}$ is reached in the region just behind the streamer head and this is lower than the background electric field. We can estimate how the choice $E_{red, min} = 0$ affects the results by writing equation~(\ref{production_ratio}) as

\begin{equation}
S_{12}  \simeq \frac{P_{1}(1-\gamma_1)}{P_{2}(1-\gamma_2)}, \label{production_ratio_I}
\end{equation}

where we define

\begin{equation}
P_{i}  = \int_{0}^{E_{red, max}} d \varepsilon  \left( E_{red, max} - \varepsilon \right)^{\beta-1} k_i\left( \varepsilon \right), \label{Ii}
\end{equation}

and

\begin{equation}
\gamma_{i}  = \frac{\int_{0}^{E_{red, min}} d \varepsilon  \left( E_{red, max} - \varepsilon \right)^{\beta-1} k_i\left( \varepsilon \right)}  {P_i}, \label{gammai}
\end{equation}

and calculating the value of the ratio of $(1 - \gamma_1)$ to $(1 - \gamma_2)$ using the streamer model at two different pressures, polarities and background electric fields. The value of this ratio ranges from 2.6 (negative laboratory streamer at atmospheric pressure with a background electric field of -15~kV~cm$^{-1}$) to 2.9 (positive sprite streamer with a background electric field of -120~V~m$^{-1}$). Therefore, the assumption that $E_{red, min}=0$ in equation~(\ref{production_ratio}) for the  diagnostic of streamer heads can introduce an error in the estimated production ratio of a factor 3 according to equation~(\ref{production_ratio_I}). This error results in a uncertainty of a 25\% over the estimated peak electric field.



\subsection{Peak electric field under the assumption of planar geometry}
\label{sec:Nmethod}

\cite{Lagarkov1994/Book} (1994, p. 62) derived a relation between the electric field and the ionization level in  a flat ionization front. \cite{Li2007/JAP} employed this relation to calculate the ionization of air molecules in streamer heads. \cite{Li2007/JAP} and \cite{Dubrovin2014/ICA} derived a relation between the electric field and the ionization level in a planar ionization front. In this section, we extend the results of \cite{Li2007/JAP} and \cite{Dubrovin2014/ICA} to show that there is also a relation between the molecular excitation level and the electric field. The resulting relation can be used to estimate the peak electric field in ionization fronts.

The characteristic time of dielectric screening in a non-equilibrium plasma can be written as

\begin{equation}
\tau_d = \frac{\epsilon_0}{\sigma}, \label{maxwelltime}
\end{equation}
where $\epsilon_0$ is the vacuum permittivity, $\sigma$ stands for the electrical conductivity which we assume is dominated by electrons. This conductivity is determined by the product of the elementary charge ($e$), the electron mobility ($\mu_e$) and the electron density ($n_e$).

Assuming a planar geometry and an external electric field that varies slowly \citep{Li2007/JAP, Dubrovin2014/ICA}, the
local electric field ($\varepsilon$) evolves as

\begin{equation}
\frac{d\varepsilon}{dt} = -\frac{e \mu_e n_e}{\epsilon_0}\varepsilon = -\frac{\varepsilon}{\tau_d}. \label{dedt}
\end{equation}

In a non-equilibrium plasma, the production rate of electronically excited molecules $N_i$ by electron impact is given by

\begin{equation}
\nu_i = \frac{dN_i}{dt} = N k_i(\varepsilon) n_e, \label{dNsdt}
\end{equation}
where $k_i(\varepsilon)$ is the reaction rate coefficient and we have neglected other processes that affect $N_i$.

The relation between the density of excited molecules and the electric field \citep{Lagarkov1994/Book} can be obtained from equations~(\ref{dedt}) and (\ref{dNsdt}),

\begin{equation}
N_i = \int_{E_{red, min}}^{E_{red, max}} d\varepsilon \frac{N \epsilon_0}{e \mu_e} \frac{k_i(\varepsilon)}{\varepsilon}. \label{dNsdte}
\end{equation}
On the other hand, the ratio between the density of two electronically excited molecules ($N_{12}$ = $N_1$/$N_2$) deduced from the recorded optical intensity and assuming that the electron mobility is not electric field dependent can be written as

\begin{equation}
N_{12} \simeq \frac{ \int_{E_{red, min}}^{E_{red, max}}  \varepsilon^{-1} k_1\left( \varepsilon \right) d \varepsilon} {\int_{E_{red, min}}^{E_{red, max}}  \varepsilon^{-1} k_2\left( \varepsilon \right) d \varepsilon }. \label{production_ratio_N}
\end{equation}

Equation~(\ref{production_ratio_N}) is an alternative to equation~(\ref{production_ratio}). Equation~(\ref{production_ratio_N}) does not depend on any particular parameter. However, using equation~(\ref{production_ratio_N}) to derive the peak electric field in the discharge also requires knowing the minimum electric field in the region where the optical emissions are produced. We can use equations~(\ref{production_ratio_I}), (\ref{Ii}) and (\ref{gammai}) together with equation~(\ref{production_ratio_N}) to evaluate the uncertainty in the estimated peak electric field under the assumption $E_{red, min}=0$. The value of the ratio $(1 - \gamma_1)$ to $(1 - \gamma_2)$ ranges between $1+\SI{3e-5}{}$ (positive laboratory streamer with a background electric field of $\SI{-12.5}{kV/cm}$) and $1+\SI{7e-4}{}$ (positive sprite streamer with a background electric field of $\SI{120}{V/m}$)). Hence, the assumption $E_{red, min}=0$ in equation~(\ref{production_ratio_N}) does not introduce a significant error in this diagnostic method. In the same manner, a possible estimation of the minimum electric field in the streamer channel does not improve this method.

\section{Comparison of methods and discussion}
\label{sec:comparison}

In this section we compare the two methods described sections~\ref{sec:Nmethod} and \ref{sec:method}. Following the notation by \cite{Celestin2010/GeoRL}, we define $\Gamma_E$ as the ratio between the peak electric field in the streamer simulation and the peak electric field derived from optical diagnostic methods.

\begin{figure}
\centering
\includegraphics[width=18cm]{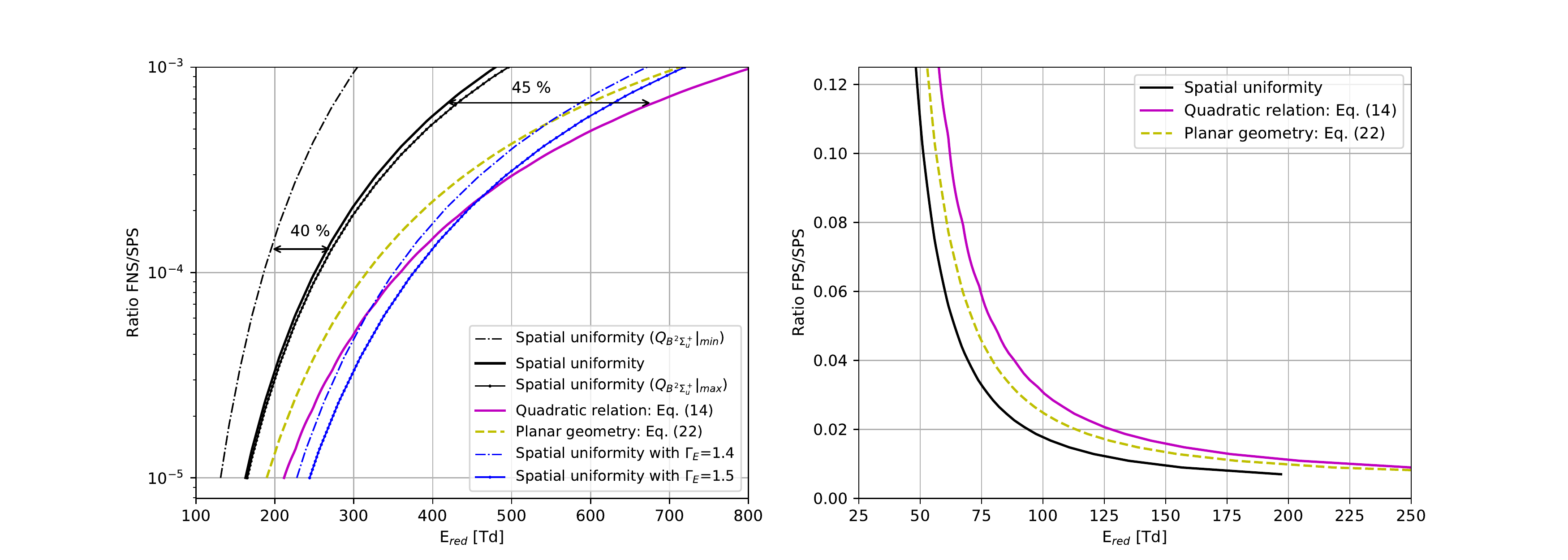}
\caption{Reduced electric field dependence of the ratios FNS$_{(0,v^{\prime\prime})}$/SPS (left panel) and FPS/SPS (right panel) at atmospheric pressure assuming 1) spatial uniformity of the electric field distribution according to equation~(\ref{equality}) (black solid line), 2) spatial non-uniformity of the electric field distribution according to equation~(\ref{production_ratio}) (purple solid line), 3) planar geometry of the electric field according to equation~(\ref{production_ratio_N}) (yellow dashed line) and 4) spatial uniformity of the electric field distribution with $\Gamma_E$=1.4 and $\Gamma_E$=1.5 \citep{Celestin2010/GeoRL} (blue lines). The employed reaction rate coefficients are collected in table~\ref{table:op_reac}.}
\label{fig:R_en}
\end{figure}

Let us first investigate the dependence between the peak electric field and the considered emission ratios. In figure~\ref{fig:R_en} we plot the peak electric field dependence of the emission ratios FNS$_{(0,v^{\prime\prime})}$/SPS and FPS/SPS using the described methods and the reaction rate coefficients of table~\ref{table:op_reac}. We have also calculated the peak electric field dependence of the emission ratio FNS$_{(0,v^{\prime\prime})}$/SPS using the lowest and highest quenching rates of N$_2$$^+$(B$^2$$\Sigma_u^+$) by air provided by \cite{bilek2018electric}. Comparison between the averaged uncertainty due to the quenching rates (40~\%) and the spatial non-uniformity (45~\%) in figure~\ref{fig:R_en} shows that the effect of the spatial non-uniformity of the discharge in the estimation of the peak electric field from the FNS$_{(0,v^{\prime\prime})}$ to SPS ratio is of the same order to that of quenching. The right panel of figure~\ref{fig:R_en} indicates that the ratio FPS/SPS cannot provide accurate information about the peak electric field above $\sim$ 200~Td.

\begin{figure}
\centering
\includegraphics[width=8cm]{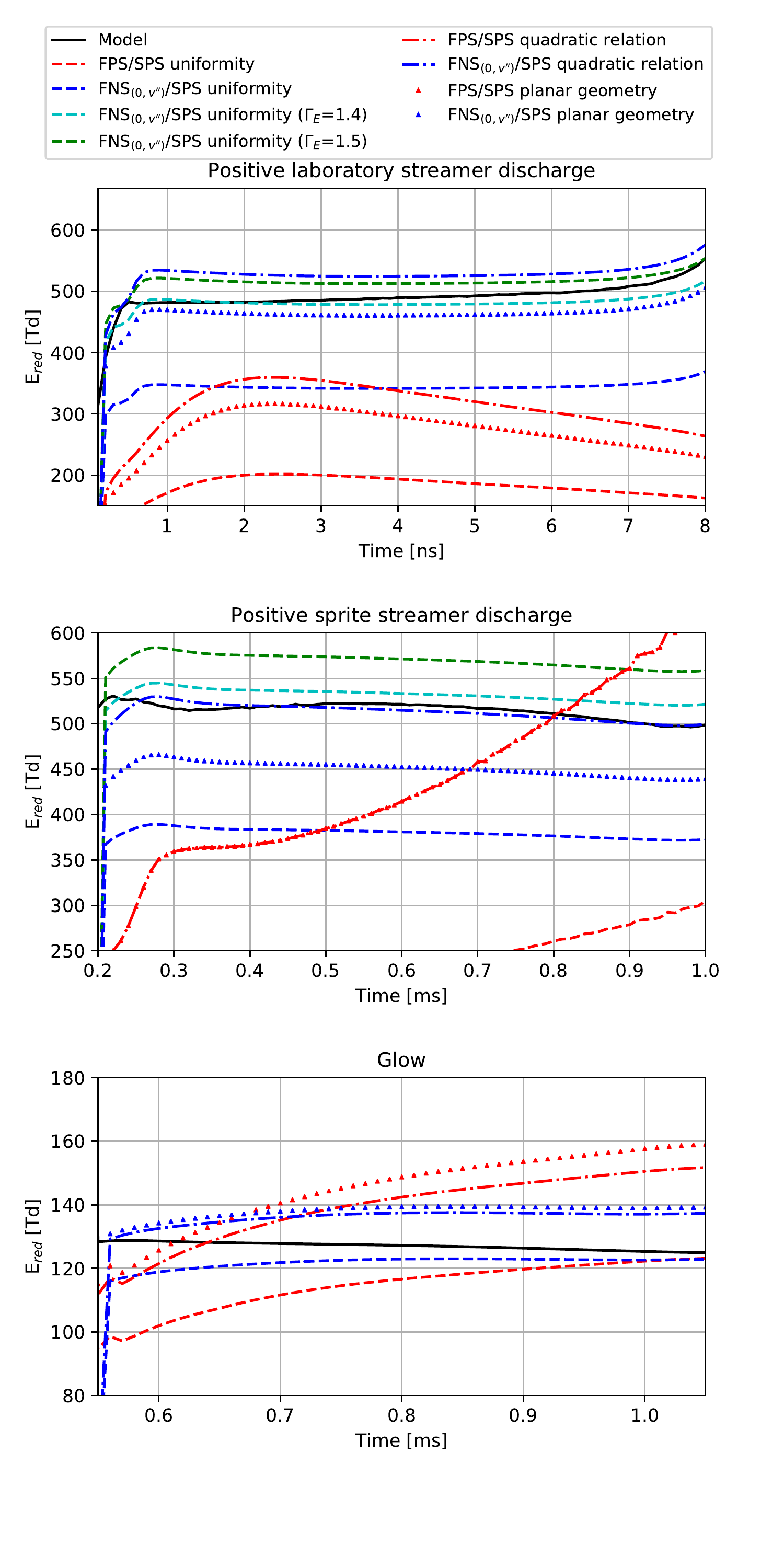}
\caption{Temporal evolution of the electric field peak in the head of two computationally simulated streamers at different pressures and in a glow discharge. Lines are as follow: Peak electric field given by the model (black solid color lines), deduced peak electric field considering that the electric field is homogeneously distributed in space (dashed color lines), deduced peak electric field considering non-uniformity of the electric field (dashed doted color lines), deduced peak electric field considering planar geometry of the electric field (triangle color lines). The non-plotted peak electric fields obtained by the ratio of FNS$_{(1,v^{\prime\prime})}$ to SPS are similar to the plotted peak electric fields obtained by the ratio of FNS$_{(0,v^{\prime\prime})}$ to SPS.}
\label{fig:ered}
\end{figure}

Figure~\ref{fig:ered} shows the result of applying the peak electric field estimation methods to a streamer head simulated at atmospheric pressure, a streamer head at low pressure (71 to 75~km) and a glow segment inside a streamer channel at low pressure. Dashed lines correspond to the peak electric field considering that the electric field is homogeneously distributed in space. Dotted and dashed dotted color lines are the electric field peak considering that the electric field is inhomogeneously distributed in space.

Considering that the electric field is inhomogeneously distributed in space in streamer heads is clearly justified. The ``quadratic relation method" and the ``planar geometry method" improve the estimation of the peak electric field with respect to the ``uniformity method" for streamer discharges. In addition, the ``quadratic relation method" improves the estimation of the peak electric field with respect to the ``uniformity method" using correction factor for the positive sprite streamer discharge.

However, doing the same to study the glow introduces more error than assuming an electric field homogeneously distributed. Figure~\ref{fig:ered} also shows that the ratio of FPS to SPS does not provide enough information about the electric field in the streamer head, where the electric field is above 200~Td.

\begin{figure}
\centering
\includegraphics[width=18cm]{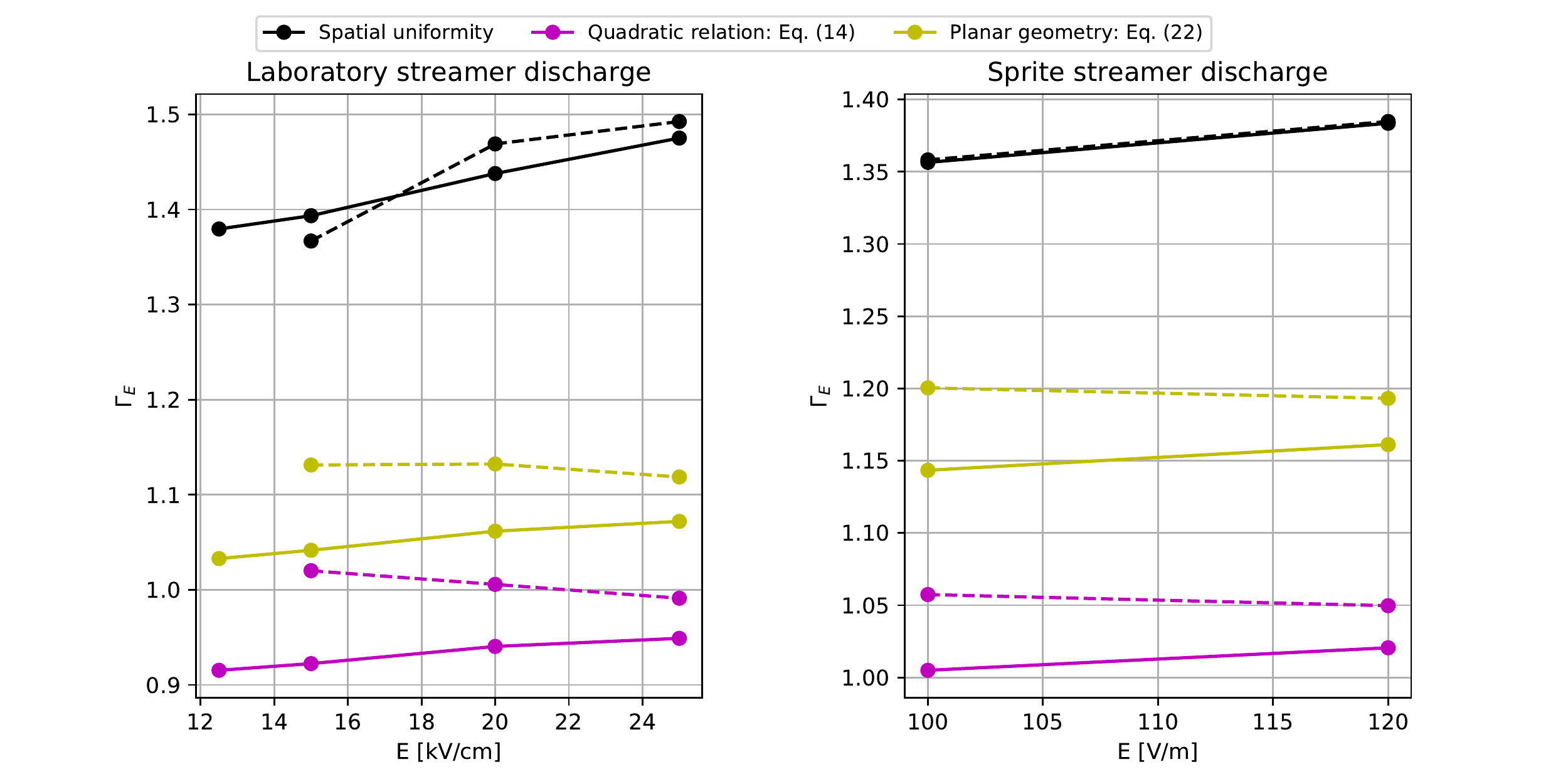}
\caption{Ratio $\Gamma_E$ between the peak electric field in the streamer simulation and the peak electric field estimated using different diagnostic methods and the emissions coefficient FNS$_{(0,v^{\prime\prime})}$/SPS. Dashed and solid lines corresponds to negative and positive streamers, respectively. Please, note that in the sprite case (right plot) \SI{100}{V/m} corresponds to \SI{120}{Td} at \SI{74.23}{km} and \SI{120}{V/m} corresponds to \SI{120}{Td} at \SI{72.91}{km}.}
\label{fig:comp_methods}
\end{figure}

In figure~\ref{fig:comp_methods} we plot the time averaged $\Gamma_E$ coefficient for the considered diagnostic methods in streamer heads. $\Gamma_E$ ranges between $\sim$ 1.4 and 1.5 and increases with the background electric field for the method based on the uniformity of the electric field, as previously reported by \cite{Celestin2010/GeoRL}. On the other hand, $\Gamma_E$ ranges between  $\sim$ 0.9 and 1.2 and has a weaker dependence on the background electric field for the methods that introduce the non-uniformity of the electric field.

Following \cite{Paris2005/JPhD}, we derive empirical formulae for the relationship between the intensity ratio FNS/SPS in streamer discharges and the peak reduced electric field using the quadratic diagnostic method based on the relationship between the electron and the electric field spatial distributions:

\begin{equation}
R_{FNS/SPS} = \exp\left[ a \left(-b \left(E_{red}\right)^{c} - d  \left(E_{red}\right)^{e}\right) \right], \label{fitstreamers}
\end{equation}

where the coefficients $a$, $b$, $c$, $d$ and $e$ are provided for different streamers discharges and emissions lines in table~\ref{table:coeff}. The electric field dependence of R$_{391.4/337}$, R$_{391.4/399.8}$, R$_{427.8/337}$ and R$_{427.8/337}$ are calculated taken the quenching rate constants, the radiative decay constants and the cross sections of electronic excitation of N$_2$$(C^{3}\Pi_u, v=0)$ and N$_2$$(C^{3}\Pi_u, v=1)$ from \cite{Gordillo-Vazquez2010/JGRA} and \cite{PerezInvernon2018a/JGR}.  The employed quenching rate constants, radiative decay constants and cross sections of electronic ionization and excitation of N$_2$$^+$$(B^2\Sigma_u^+, v=0)$ and N$_2$$^+$$(B^2\Sigma_u^+, v=1)$ are listed in table~\ref{table:op_reac}.

\small
\begin{table}
\caption{Coefficients $a$, $b$, $c$, $d$ and $e$ of equation~(\ref{fitstreamers}) for different streamers discharges and emissions lines.} \label{table:coeff}
\scalebox{0.7}{
\begin{tabular}{l*{3}{c}r}
\hline
       Type of streamer   & Ratio of lines & $a$, $b$, $c$, $d$, $e$  \\
\hline
Laboratory streamer & R$_{391.4/337}$  & 2.22$\times$10$^{2}$, 0.41, -0.39, 3.93$\times$10$^{2}$, -1.89 \\
 & R$_{427.8/337}$  & 32.88,  60.24, -0.98,  0.15,   -4.10 $\times$10$^{-4}$\\
& R$_{427.8/399.8}$  & 1.93$\times$10$^{3}$, 3.86$\times$10$^{3}$, -2.82,  7.89$\times$10$^{-2}$, -0.46 \\
 & R$_{391.4/399.8}$  & 1.67$\times$10$^{2}$,  0.15, -0.24, 46.44, -1.31\\
Sprite streamer & R$_{391.4/337}$  & 1.01$\times$10$^{5}$, 9.74$\times$10$^{-3}$, -0.84, 2.02$\times$10$^{5}$, -4.59\\
& R$_{427.8/337}$  & 1.54$\times$10$^5$, 3.08$\times$10$^{5}$, -4.63, 3.92$\times$10$^{-3}$, -0.74 \\
& R$_{427.8/399.8}$  & 3.45$\times$10$^{3}$, 6.90$\times$10$^{3}$, -3.16,  0.52, -0.72 \\
& R$_{391.4/399.8}$  & 82.37,  9.97$\times$10$^7$, -4.08$\times$10$^5$,  1.68,  -0.90 \\
\end{tabular}
}
\end {table}
\normalsize



\section{Application of the described methods to experimental data}
\label{sec:signal_armostrong}

\cite{Armstrong1998/JASTP} reported the ratio of SPS$_{(1,4)}$ (399.8~nm) to FNS$_{(1,0)}$ (427.8~nm) emitted by TLEs detected from the FMA Research Yucca Ridge Field station, located at an altitude of about 1500~m. The photometric measurements in the SPRITE's 95 and 96 campaigns reported by \cite{Armstrong1998/JASTP} were recorded using two photometers with a time resolution of 1.3~ms. \cite{Armstrong1998/JASTP} discussed the overlap between the SPS and the FNS bands in the 427.8~nm photometer due to the wavelengths dependence of the response of the employed instruments. According to their estimation, the signal of the SPS in the 427.8~nm photometer corresponds to approximately 26\% of the signal in the 399.8~nm photometer. Therefore, 26\% of the signal reported by the 399.8~nm photometer has to be subtracted from the signal recorded by the 427.8~nm photometer to get the true contribution of the FNS.

In this section, we estimate the peak electric field using the reported intensity ratio of two sprites referred as ``DAY 201 - 19050651" and ``DAY 201 - 19062205" in \cite{Armstrong1998/JASTP}. The Field of View (FOV) of the photometers covers a large spatial region of the sprites (see in figures~5 and 6 of \cite{Armstrong1998/JASTP}). Thus, the reported intensity ratio is a combination of the optical emissions of the streamers and glows that form the sprite. Unfortunately, we cannot separate optical emissions from streamers and glows to decide the convenience of considering that the electric field is homogeneously or inhomogeneously distributed in space. Then, we calculate the peak electric field from the reported ratio using both methods. We follow the same procedure:

\begin{enumerate}
\item The reported ratios of SPS$_{(1,4)}$ (399.8~nm) to FNS$_{(1,0)}$ (427.8~nm) for the sprites ``DAY 201 - 19050651" and ``DAY 201 - 19062205" in \cite{Armstrong1998/JASTP} are, respectively, 1.91 and 2.84 when the maximum luminosity is reached. We correct these ratios by considering that the signal of the SPS in the 427.8~nm photometer corresponds to approximately 26\% of the signal in the 399.8~nm photometer \citep{Armstrong1998/JASTP}.
\item The observed ratio of intensities is influenced by the atmospheric transmittance. The sprites were triggered by a storm which was located $\sim$260~km away from the observer \citep{Armstrong1998/JASTP}, while the observatory station altitude is $\sim$1500~m above sea level. In addition, we assume that the optical emissions are produced at $\sim$70~km, which is a characteristic altitude of sprites \citep{Stenbaek-Nielsen2010/JGRA, luque2016sprite}. Therefore we can calculate the optical transmittance of the atmosphere between the sprites and the photometers using the software MODTRANS~5 \citep{berk2005modtran}. We use the calculated optical transmittance to obtain the emitted ratio of intensities from the recorded signal.
\item We calculate the production of emitting molecules by electron impact using equations~(\ref{densities}) and (\ref{productionc}) and assuming the air density at an altitude of 70~km. We use the reaction rate coefficients of table~\ref{table:op_reac} to calculate the density of N$_2$$^+$$(B^2\Sigma_u^+, v=1)$. The vibrational kinetics employed to calculate the density of N$_2$$(C^{3}\Pi_u, v^{\prime}=1)$ is taken from \cite{Gordillo-Vazquez2010/JGRA} and \cite{PerezInvernon2018a/JGR}.
\item We estimate the peak electric field in the sprites by considering a homogeneously (inhomogeneously) distributed electric field in space according to equation~(\ref{equality}) (equations~(\ref{production_ratio}) or (\ref{production_ratio_N})) with $E_{red, min}$ equal to zero.

\end{enumerate}

The resulting peak electric field for the sprite ``DAY 201 - 19050651" if we assume a homogeneously distributed electric field in space is 450~Td, while for the sprite ``DAY 201 - 19062205" the peak electric field is 326~Td. Using the previously defined quadratic method (equations~(\ref{production_ratio})), the resulting peak electric field for the sprite ``DAY 201 - 19050651" is 757~Td), while for the sprite ``DAY 201 - 19062205" the peak electric field is 503~Td. The use of equation~(\ref{production_ratio_N}) would lead to slightly lower reduced electric fields, as seen in figure~\ref{fig:R_en}. As photometers cannot spatially resolve the emissions, we are probably analyzing combined optical emissions from streamer heads and glows. Therefore, we cannot determine which method is the most accurate in this case. As the reported intensities are a combination of the intensities emitted by streamers and glows, we can assume that the value of the peak electric fields in streamer heads of the sprites ``DAY 201 - 19050651" and ``DAY 201 - 19062205" are respectively in the range 451~Td - 757~Td and 327~Td - 503~Td. These derived values are probably influenced by the peak electric field inside glows (on the order of 120~Td \citep{luque2016sprite}) and streamer heads (several hundreds of Td).
We have repeated these calculations for the case of a sprite altitude of 80~km instead of 70~km, obtaining an increase in the peak electric fields of 1.5\%.

\section{Conclusions}
\label{sec:conclusions}

We have used a streamer model to simulate streamer heads and glows to quantify the influence of the non-uniformity of the electric field in spectroscopic diagnostic methods. The analysis of the spatial inhomogeneity of the electric field in air discharges has allowed us to improve the optical diagnostic methods commonly employed in the determination of the peak electric field in streamer heads. The commonly employed method underestimates the peak electric field by about 40\%-50\%, while the methods developed in this work reduce the uncertainty to about 10\%-20\%. We have also showed that the ratio of FPS to SPS can be employed to deduce the peak electric field in streamer glows without considering the spatial inhomogeneity of the electric field.

The first developed optical diagnostic method (section~\ref{sec:method}) is based on the characterization of the non-uniformity of the electric field in streamer heads using a streamer model. This method introduces an exponent ($\beta$ $\simeq$ 2) that is almost constant for different streamer configurations. The most important uncertainty in the peak electric field calculated with this method is due to the uncertainty in the electric field inside the streamer channel. In general, the value of $E_{red, min}$ is unkown.  Hence, we propose to set $E_{red, min}=0$.

The second developed optical diagnostic method (section~\ref{sec:Nmethod}) is based on the relation between the electric field and the level of molecular excitation in a planar ionization front. This method does not introduce any extra parameter to estimate the peak electric field and considering $E_{red, min}=0$ does not introduce a significant error. Thus, in general, it is more convenient than the method described in section~\ref{sec:method}. However, the method described in section~\ref{sec:method} can be useful whenever the estimation of the electric field in the streamer channel is possible. In principle both methods can be generalized to other gases if the appropriate emission lines are identified.

Despite the improvements, optical diagnostic methods of air discharges from the ratio of FNS to SPS at atmospheric pressure are still very sensitive to the considered chemical reactions rates \citep{obrusnik2018electric, bilek2018electric}. As \cite{obrusnik2018electric} and \cite{bilek2018electric} concluded, more efforts are needed for a more precise determination of the reaction rates (especially quenching rates) that are important for diagnostic methods based on the FNS emission at atmospheric pressure.

The uncertainty in the reaction rates employed in the determination of the peak electric field from the ratio of FPS to SPS is lower than the uncertainty in the reaction rates involved in the FNS emissions. Nevertheless, the FPS/SPS ratio of intensities is only applicable for glow discharges where the electric field is known to be below $\sim$200~Td.

\begin{acknowledgments}
This work was supported by the Spanish Ministry of Science and Innovation, MINECO under project ESP2017-86263-C4-4-R and by the EU through the European Research Council (ERC) under the European Union's H2020 programm/ERC grant agreement 681257. This project has received funding from the European Union's Horizon 2020 research and innovation programme under the Marie Sk\l{}odowska-Curie grant agreement SAINT 722337. The authors acknowledge financial support from the State Agency for Research of the Spanish MCIU through the ``Center of Excellence Severo Ochoa" award for the Instituto de Astrof\'isica de Andaluc\'ia (SEV-2017-0709). FJPI acknowledges a PhD research contract, code BES-2014-069567. Data and codes used to generate figures presented here are available at \url{https://cloud.iaa.csic.es/public.php?service=files&t=54fcc1134f55efa61087670678226236}.
\end{acknowledgments}

\newcommand{\jcp}{J. Chem. Phys. }
\newcommand{\ssr}{Space Sci. Rev.}
\newcommand{\planss}{Plan. Spac. Sci.}
\newcommand{\pre}{Phys. Rev. E}
\newcommand{\nat}{Nature}
\newcommand{\icarus}{Icarus}
\newcommand{\ndash}{-}

\bibliographystyle{agu08}

\end{article}

\end{document}